\documentclass{elsart}

 \usepackage{graphicx}

\begin{document}

\begin{frontmatter}



\title{Branched-polymer to inflated transition of self-avoiding fluid surfaces}


\author{Hiroshi Koibuchi}
\ead{koibuchi@mech.ibaraki-ct.ac.jp}

\address{Department of Mechanical and Systems Engineering, Ibaraki National College of Technology, 
Nakane 866, Hitachinaka,  Ibaraki 312-8508, Japan}

\author{Andrey Shobukhov}

\address{Faculty of Computational Mathematics and Cybernetics, Lomonosov Moscow State University, 119991, Moscow, Leninskiye Gory, MSU, 2-nd Educational Building, Russia}

\begin{abstract}
We study phase transition of self-avoiding fluid surface model on dynamically triangulated lattices using the Monte Carlo simulation technique. We report the continuous transition between the branched polymer and inflated phases at ${\it \Delta}p \!=\!0$, where  ${\it \Delta}p (=\!p_{\rm in}\!-\!p_{\rm out})$ is the pressure difference between the inner and outer sides of the surface. This transition is characterized by almost discontinuous change of the enclosed volume versus the variations of the bending rigidity $\kappa$ and the pressure difference ${\it \Delta}p$. No surface fluctuation transition accompanies this transition up to the surface with the number of vertices $N\!=\!2562$.
\end{abstract}

\end{frontmatter}


\section{Introduction}
 A good example of the fluid surface is the so-called bi-continuous oil-water interface.  Random  bi-continuous structure is expected to be separated from stable (for exapmple, lamellar) structure by  phase transition at finite curvature elasticity $\kappa_c$ \cite{deGennes-Taupin-JPC1982,Safran-LANGMUIR1991}. For $\kappa \!<\!\kappa_c (\kappa \!>\!\kappa_c)$, the entropy (curvature) effect is dominant, and therefore the random bi-continuous (lamellar) structure is more likely to appear. The stability of these different phases is established with the help of the Helfrich Hamiltonian for membranes. 

However, this transition is still unclear for triangulated surfaces. This is in sharp contrast to the crumpling transition in the surface models of Helfrich and Polyakov for polymerized or tethered membranes, which have  been studied theoretically and numerically for a long time on the basis of the statistical mechanics \cite{HELFRICH-1973,POLYAKOV-NPB1986,KLEINERT-PLB1986,WHEATER-JP1994,NELSON-SMMS2004,KANTOR-NELSON-PRA1987,Bowick-PREP2001,GOMPPER-KROLL-SMMS2004}. The basic property of the model is that the surface becomes smooth at sufficiently large bending rigidity $\kappa$, while it collapses at $\kappa\to 0$ \cite{P-L-1985PRL,PKN-PRL1988,DavidGuitter-1988EPL,Kownacki-Mouhanna-2009PRE,NISHIYAMA-PRE-2004-2010}. The collapsed phase of the triangulated surface models is known to be strongly dependent on the self-avoiding (SA) interaction \cite{KANTOR-KARDAR-NELSON-PRL1986,KANTOR-KARDAR-NELSON-PRA1987}. In fact, no collapsed phase appears on connection-fixed surfaces at $\kappa\to 0$ under the zero pressure difference  ${\it \Delta}p\!:=\!p_{\rm in}\!-\!p_{\rm out}\!=\!0 $, where $\!p_{\rm in}(p_{\rm out})$ is the pressure inside (outside) the surface \cite{PLISCHKE-BOAL-PRA1988,Ho-Baum-EPL1990,Grest-JPhysI1991,Kroll-Gompper-JPF1993,Munkel-Heermann-PRL1995,BCTT-PRL2001,BOWICK-TRAVESSET-EPJE2001}.  

As far as the fluid vesicles are concerned, the phase structure of a self-avoiding model for them with varying the bending rigidity $\kappa$ and the pressure difference  ${\it \Delta}p$ was numerically studied in \cite{Gompper-Kroll-PRA1992,Gompper-Kroll-EPL1992,Gompper-Kroll-PRE1995,Dammann-etal-JPIF1994}. Free diffusion of lipids in fluid vesicles was simulated by the diffusion of vertices on dynamically triangulated random lattices. At present it is well known from the model of Gompper-Kroll that the inflated and stomatocyte phases are separated by a first-order transition at negative pressure difference (${\it \Delta}p \!<\! 0)$, and that the inflated and branched polymer (BP) phases are also separated by a first-order transition at the same condition for ${\it \Delta}p$. However, it is still unclear whether the BP phase is separated from the inflated phase by a phase transition at ${\it \Delta}p\!\to\!0$.

In this paper, we numerically study the surface model of Helfrich and Polyakov on dynamically triangulated SA lattices of sphere topology, which is almost identical with the model of Gompper-Kroll for fluid vesicles. Using the notion of curvature elasticity, we demonstrate exponential behavior of the persistence length $\xi$ with respect to the bending rigidity at the transition point. We focus on the phase transition between the inflated and BP phases at the zero pressure difference  ${\it \Delta}p \!\simeq\!0$. 

The main difference between the model in this paper and the model of Gompper-Kroll in \cite{Gompper-Kroll-PRE1995} is in the SA interaction. The model of Ref. \cite{Gompper-Kroll-PRE1995} is the so-called  beads-spring model, which consists of hard spheres of diameter $\sigma_0$ connected by flexible tethers (or bonds) of length $\ell_0 <\sqrt{3}\sigma_0$. This constraint for the bond length protects the beads from penetrating the triangles and makes the surface self-avoiding. To the contrary, the SA interaction included in the Hamiltonian of the model in this paper prohibits two disconnected (or disjoint) triangles from intersecting with each other \cite{Kroll-Gompper-JPF1993,BOWICK-TRAVESSET-EPJE2001}.  

\section{Model}
The triangulated sphere is obtained from the icosahedron by splitting its edges and faces. The total number $N$ of vertices equals $N\!=\!10\ell^2\!+\!2$, where $\ell$ is the number of bond partitions. The total number of bonds $N_B$, and the total number of triangles $N_T$ are are equal to $N_B\!=\!30\ell^2$ and $N_T\!=\!20\ell^2$ respectively. The numbers $N, N_B, N_T$ remain unchanged during the dynamical triangulation, which will be described below. The vertex coordination number $q$ equals $6$ except for the $12$ vertices of the initial icosahedron; and for them $q$ equals $5$.

\begin{figure}[htb]
\centering
\includegraphics[width=11cm]{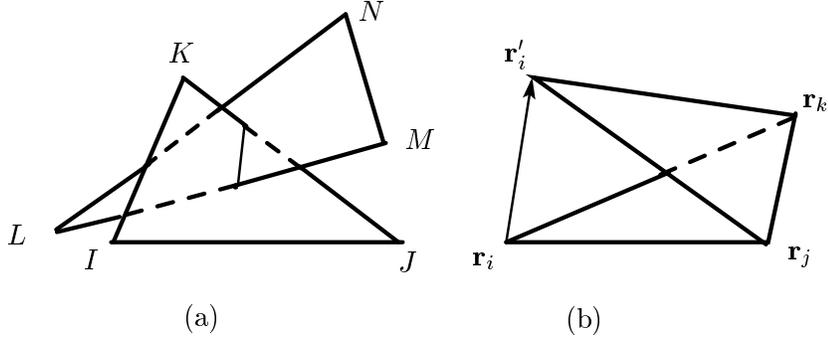}
\caption{ (a) Two disjoint and intersecting triangles ${\it \Delta}_{IJK}$ and ${\it \Delta}_{LMN}$, (b) a new vertex position ${\bf r}_i^\prime$ of ${\bf r}_i$ as a MC update for a vertex move, and the corresponding triangles ${\it \Delta}_{ijk}$ and ${\it \Delta}_{i^\prime jk}$, where $i^\prime$ denotes ${\bf r}_i^\prime$.} 
\label{fig-1}
\end{figure}

The Hamiltonian of the model is  a linear combination of the Gaussian bond potential $S_1$, the bending energy $S_2$, the pressure term $-{\it \Delta p}V$, and the SA potential $U$, such that:
\begin{eqnarray}
\label{Disc-Eneg} 
&& S({\bf r}, {\mathcal T})=S_1 + \kappa S_2 - {\it \Delta p}V + U,  \\
&& S_1=\sum_{ij} \left( {\bf r}_i-{\bf r}_j\right)^2, \quad S_2=\sum_{(ij)} (1-{\bf n}_i \cdot {\bf n}_j), \nonumber
\end{eqnarray} 
where $S({\bf r}, {\mathcal T})$ means that the Hamiltonian depends on the vertex position ${\bf r} (\in {\bf R}^3)$ and the triangulation ${\mathcal T}$. The symbol $\kappa [kT]$ denotes the bending rigidity. $\sum_{ij}$ in $S_1$ represents the sum over all bonds $ij$, ${\bf n}_i$ in $S_2$ is a unit normal vector of the triangle $i$, and $\sum_{(ij)}$ in $S_2$ is the sum over all nearest neighbor triangles $i$ and $j$. The SA potential $U$ looks as follows:
\begin{eqnarray} 
\label{SA-potential}
&& U=\sum_{\it \Delta \Delta^\prime} U({\it \Delta,\Delta^\prime}), \\
&& U({\it \Delta,\Delta^\prime})= \left\{
     \begin{array}{@{\,}ll}
    \infty & \; ({\rm triangles}\;{\it \Delta \Delta^\prime} \; {\rm intersect}  ), \\
             0 & \; ({\rm otherwise}), 
     \end{array} 
               \right. \nonumber \\ \nonumber
\end{eqnarray}
where $\sum_{\it \Delta \Delta^\prime} $ denotes the sum over all pairs of disjoint triangles ${\it \Delta}$ and ${\it \Delta}^\prime$. Figure \ref{fig-1}(a) shows a pair of two disjoint triangles ${\it \Delta}_{IJK}$ and ${\it \Delta}_{LMN}$, that intersect with each other, and hence $U({\it \Delta}_{IJK},{\it \Delta}_{LMN})\!=\!\infty$.

The SA interaction given by Eq. (\ref{SA-potential}) is identical to the one assumed in Ref. \cite{Kroll-Gompper-JPF1993} and different from that of the ball spring model in Ref. \cite{KANTOR-KARDAR-NELSON-PRA1987}, where a finite size of ball and a constrained bond length prohibit the balls from penetrating the triangles as mentioned in the final part of the introduction. The SA interaction in this paper is also different from the impenetrable plaquette model in Ref. \cite{BOWICK-TRAVESSET-EPJE2001}, because the SA potential in Eq. (\ref{SA-potential}) is defined to be zero or infinite while the one in Ref. \cite{BOWICK-TRAVESSET-EPJE2001} is defined to have finite nonzero values. However, the final results are expected to be independent of the definition of the SA interaction.

The fluid surface model is defined by the partition function
\begin{equation} 
\label{Part-Func-2}
Z = \sum_{\mathcal T} \int^\prime \prod _{i=1}^{N} d {\bf r}_i \exp\left[-S({\bf r}, {\mathcal T})\right],
\end{equation} 
where  $\sum_{\mathcal T}$ denotes the sum over all possible triangulations \cite{Baum-Ho-PRA1990,CATTERALL-NPBSUP1991,AMBJORN-NPB1993}. The symbol $\int^\prime \prod _{i=1}^{N} d {\bf r}_i $ in $Z$  means that the $3N$-dimensional integrations are performed by fixing the center of mass of the surface to the origin of ${\bf R}^3$.

\section{Monte Carlo technique}
We use the canonical Metropolis algorithm to update the variables ${\bf r}$ and ${\mathcal T}$. The variable ${\bf r}_i$ of the vertex $i$ is updated such that ${\bf r}_i\to {\bf r}_i^\prime \!=\!{\bf r}_i+\delta {\bf r}$, where $\delta {\bf r}(\in {\bf R}^3)$ is a random three-dimensional vector in a sphere of radius $r_0$. We update the triangulation variable ${\mathcal T}$ of triangulations by using the so-called bond flip technique. The variable updates are accepted with the probability ${\rm Min}[1,\exp(-\delta S)]$, $\delta S\!=\!S({\rm new})\!-\!S({\rm old})$, under the constraint of the SA potential $U$.  The acceptance rate for the update of ${\bf r}$ is given by $R_U \times R_{\bf r}$, where $R_U$ is the acceptance rate for ${\bf r}^\prime$ which satisfies the constraint $U\!=\!0$, and $R_{\bf r}$ is the Metropolis acceptance rate. The radius $r_0$ of the small sphere for $\delta {\bf r}$ is fixed to some constant so that we have about $50\%$ total acceptance rate. Not only $R_{\bf r}$ but also $R_U$ depends on $\kappa$. As for $R_U$, almost all updates are accepted for sufficiently large values of $\kappa$, while we have only $70\%\sim 80\%$ acceptance rate $R_U$ for small values of $\kappa$ including $\kappa\!=\!0$. One Monte Carlo sweep (MCS) consists of $N$ updates of ${\bf r}$ and $N$ bond flips for the update of ${\mathcal T}$. In the bond flip procedure, $N$ bonds are randomly selected from the $N_B$ bonds that are consecutively numbered. 

Let us describe a discretization technique for the SA potential $U$ in Eq. (\ref{SA-potential}), which prohibits the disjoint triangles from intersecting as described above. We show in detail how to implement the self-avoidance while updating ${\bf r}$. We successfully perform this by choosing a new vertex position ${\bf r}_i^\prime$ of the triangle  ${\it \Delta}_{i^\prime j k}$, as shown in Fig. \ref{fig-1}(b), such that the following two conditions are satisfied: 
\begin{enumerate}
\item [({\romannumeral 1})] The new triangle ${\it \Delta}_{i^\prime j k}$ does not intersect with all other disjoint bonds, which are disjoint with ${\it \Delta}_{i^\prime j k}$.

\item [({\romannumeral 2})] Each of the new bonds $i^\prime j$ and  $i^\prime k$ does not intersect with all other disjoint triangles, which are disjoint with the bonds $i^\prime j$ and  $i^\prime k$. 
\end{enumerate}

It is very time consuming to check these conditions, however, the computational time can be reduced, because it is not always necessary to check all the disjoint bonds in ({\romannumeral 1}) and all the disjoint triangles in  ({\romannumeral 2}). Indeed, it is sufficient to check only bonds and triangles inside the sphere of radius $R$. So we can considerably reduce the computational time for checking the conditions   ({\romannumeral 1}) and  ({\romannumeral 2}). The radius $R$ of the sphere is fixed to the maximum bond length computed every MCS. The center of this sphere is fixed to the center of ${\it \Delta}_{i^\prime j k}$ in ({\romannumeral 1}), while it is  fixed to the center of the bond $i^\prime j$ (or $i^\prime k$) in ({\romannumeral 2}). 

 We should note also that it is straight forward to impose the conditions ({\romannumeral 1}) and ({\romannumeral 2}) on the update of ${\mathcal T}$. Indeed, we have a new bond and two new triangles in a bond flip for the update of ${\mathcal T}$, and consequently the conditions ({\romannumeral 1}) and ({\romannumeral 2}) can be imposed on the new bond and triangles likewise for the update of ${\bf r}$. It should also be checked whether the surface is completely self-avoiding. We check it every 500 MCS to see whether or not a bond and a triangle intersect with each other for all disjoint pairs of them.

The total number of MCS is $2\times 10^7\sim 3\times 10^7$ at the transition region of the $N\!=\!2562$ surface for ${\it \Delta}p\!=\!0$ after the thermalization MCS, which is approximately $5\times 10^5$. The total number of MCS is relatively small at non transition region and on the smaller surfaces. The total number of MCS at $\kappa\!=\!1.5$ is almost the same as that of the simulations under ${\it \Delta}p\!=\!0$. The coordination number $q$ is constrained to be in the range $3\leq q \leq 30$ during the simulations. Almost all $q$ are found to be smaller than $q_{\rm max}\!=\!30$ even in the limit of $\kappa\to 0$. Thus, we expect that the results are independent of the assumed value for $q_{\rm max}$.

\section{Simulation results}
\subsection{Zero pressure ${\it \Delta p}\!=\!0$}
\begin{figure}[htb]
\centering
\includegraphics[width=11cm]{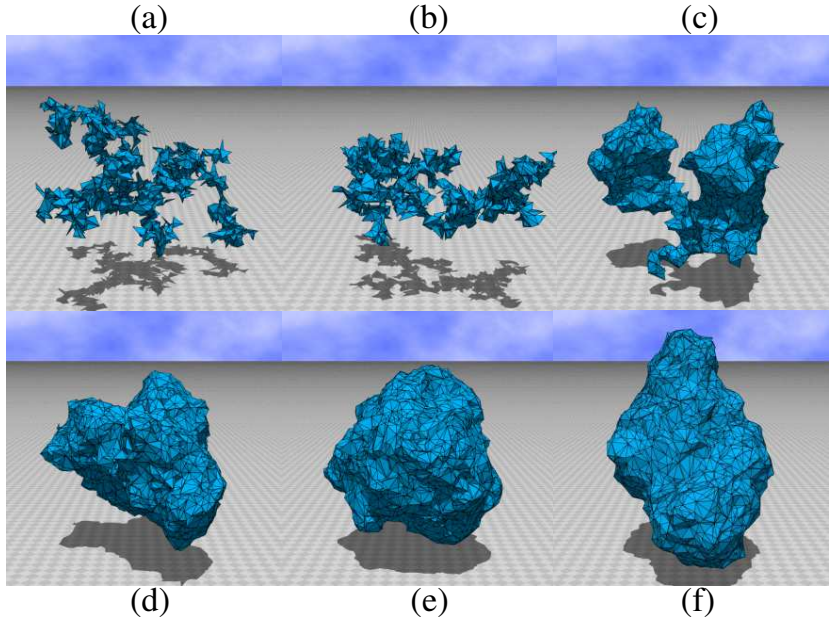}
\caption{(Color online) Snapshots of the fluid surface of size $N\!=\!2562$ with (a) $\kappa\!=\!0$, (b) $\kappa\!=\!0.1$, (c) $\kappa\!=\!1.3$, (d) $\kappa\!=\!1.5$, (e) $\kappa\!=\!1.7$, and (f) $\kappa\!=\!1.9$ for ${\it \Delta p}\!=\!0$. The snapshots are drawn in the same scale.} 
\label{fig-2}
\end{figure}
In this subsection, we present the numerical results obtained under ${\it \Delta p}\!=\!0$.  In Figs. \ref{fig-2}(a)--\ref{fig-2}(f), we show the snapshots of the surface with $N\!=\!2562$ obtained in the range $0\!\leq\!\kappa\!\leq\!1.9$. We see that the surface changes from a BP surface to an inflated one as $\kappa$ varies from $\kappa\!=\!0$ to $\kappa\!=\!1.9$. The surface is considered to be in the BP phase at $\kappa\leq 1.3$, while it is in the inflated one at $\kappa\geq 1.5$. Thus, with these snapshots we confirm that the BP phase and the inflated phase can actually be observed in the considered model at sufficiently small and large values of $\kappa$. This was actually not observed in Ref. \cite{Gompper-Kroll-PRE1995}, probably due to the small size of lattices compared to the current simulations. 

The problem that should be considered is whether or not these two different phases are separated by a phase transition, and what is the order of the transition. One can expect that the BP phase and the inflated phase are separated by a true phase transition. We shall demonstrate that this expectation is true. 

\begin{figure}[tbh]
\centering
\includegraphics[width=11cm]{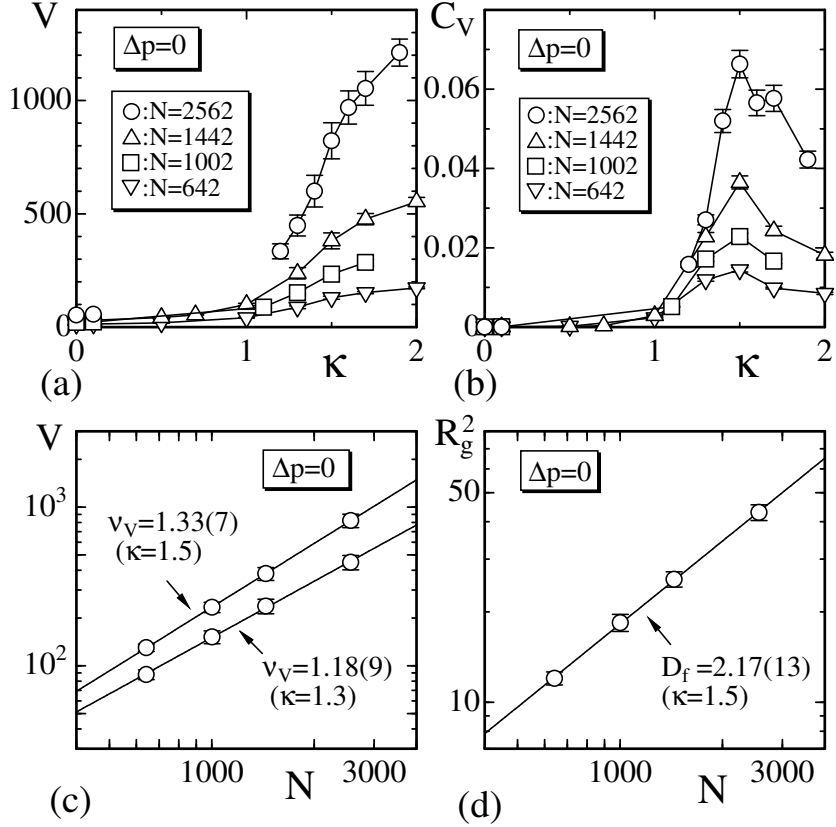}
\caption{ (a) The enclosed volume $V$ vs. $\kappa$, (b) the variance $C_{V}$ vs. $\kappa$, (c) $V$ vs. $N$ in a log-log scale, and (d) The mean square radius of gyration $R_g^2$  vs. $N$ in a log-log scale. The solid lines connecting the data symbols in (a) and (b) are drawn as a guide to the eyes.}
\label{fig-3}
\end{figure}
In Fig. \ref{fig-3}(a), we show the enclosed volume $V$ vs. $\kappa$. We see that the increase of $V$ is apparent at $\kappa\simeq 1.5$. The variance $C_V$ defined as
\begin{equation}
\label{CV}
C_{V}=(1/N^{3/2}) \langle \left(V-\langle V\rangle \right)^2 \rangle
\end{equation}
is shown in Fig. \ref{fig-3}(b) versus $\kappa$. We should explain why $N^{3/2}$ is used in the definition of $C_V$ in place of $N$ in Eq. (\ref{CV}). This is because the enclosed volume $V$ is not regarded as an intrinsic variable of the surface but is expected to vary proportional to $N^{3/2}$ at most. We can also see from Fig. \ref{fig-3}(b) that $C_V$ has an expected peak $C_V^{\rm max}$ at $\kappa\!\simeq\! 1.5$, and that $C_V^{\rm max}$ increases with increasing $N$.

The mean square radius of gyration $R_g^2$ is defined as
\begin{equation}
\label{X2}
R_g^2=(1/N) \sum_i \left({\bf r}_i-\bar {\bf r}\right)^2, \quad \bar {\bf r}=(1/N) \sum_i {\bf r}_i.
\end{equation}
From the expressions for $R_g^2$ and $V$, we obtain the size exponents $\nu_V$, ${\bar \nu}_V$ and $\nu_{R^2}$ defined by
\begin{equation}
\label{size-exponents}
V\sim N^{\nu_V}, \quad V\sim N^{(3/2){\bar \nu}_V}, \quad R_g^2\sim N^{\nu_{R^2}} =N^{2/D_f},
\end{equation}
where $\nu_V$ and ${\bar \nu}_V$ are dependent on each other in such a way that $\nu\!=\!(3/2){\bar \nu}_V$, and $D_f(=\!2/\nu_{R^2})$ is the fractal dimension (Figs. \ref{fig-3}(c), \ref{fig-3}(d)). The exponents are presented in Table \ref{table-1}. The result $\nu_V\!=\!1.18\pm0.09$ at $\kappa\!=\!1.3$ slightly deviates from $\nu_V\!=\!1$, and this deviation is compatible with the fact that the surface shown in Fig. \ref{fig-2}(c) is an object with functional dimension between one and two. Indeed,  some parts of the branched-polymer-like surface in Fig. \ref{fig-2}(c) are inflated. In fact, the volume $V$ of the true BP surface is expected to be $V\!\propto\! L$, where $L$ is the total length of the surface, and moreover, $L$ is also expected to be proportional to the surface area, which is proportional to $N$ \cite{Gompper-Kroll-PRA1992,Gompper-Kroll-EPL1992,Gompper-Kroll-PRE1995}. Thus, we have $V\!\sim\! N$ on the true BP surface and understand the reason for the deviation of  $\nu_V(=\!1.18\pm0.09)$ from $\nu_V\!=\!1$ at $\kappa\!=\!1.3$. On the other hand, we expect that $\nu_V\!=\!3/2$ in the inflated phase in the limit of $\kappa\!\to\!\infty$, where the surface becomes a sphere. Therefore, we understand the reason why the results shown in Table \ref{table-1} satisfy $1\!<\!\nu_V\!<\!3/2$ and increase with increasing $\kappa$.     
\begin{table}[hbt]
\caption{The size exponents $\nu_V$, ${\bar \nu}_V$, and $\nu_{R^2}(=2/D_f)$ in Eq. (\ref{size-exponents}) obtained for ${\it \Delta p}\!=\!0$.}
\label{table-1}
\begin{center}
 \begin{tabular}{ccccccccc}
 \hline
 $\kappa$ && $\nu_V$ && ${\bar \nu}_V$ && $\nu_{R^2}$ && $D_f$    \\
 \hline
 $1.3$ && $1.18\pm0.09$ && $0.79\pm0.06$ && $0.91\pm0.12$ && $2.20\pm0.30$ \\
 $1.5$ && $1.33\pm0.07$ && $0.89\pm0.05$ && $0.92\pm0.05$ && $2.17\pm0.13$ \\
 $1.7$ && $1.40\pm0.05$ && $0.93\pm0.03$ && $0.98\pm0.06$ && $2.03\pm0.11$ \\
 \hline
 \end{tabular}
\end{center}
\end{table}

We calculate the fractal dimension $D_f$ in the limit of $\kappa\to 0$ in order to check that the simulations are correctly performed. It is expected that $D_f\!\to\! 2 (\Leftrightarrow \nu_{R^2}\!=\!1)$ in the BP phase in the limit of $\kappa\!\to\! 0$ as described above.

\begin{figure}[htb]
\centering
\includegraphics[width=11cm]{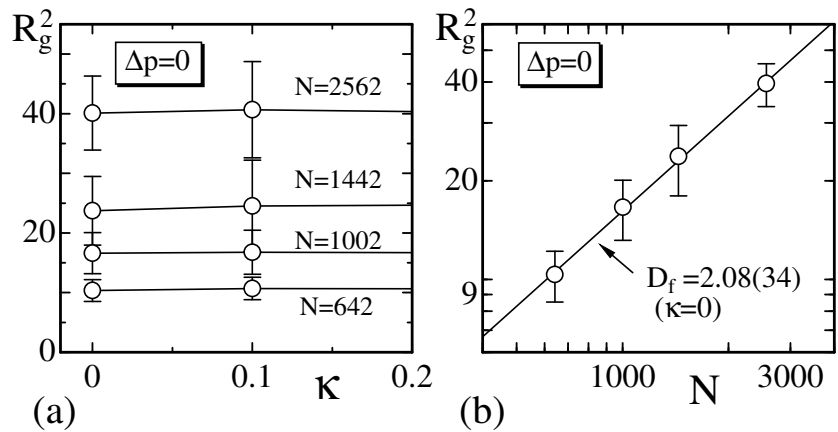}
\caption{ (a) The mean square radius of gyration $R_g^2$ vs. $\kappa$ for $\kappa\simeq 0$, (b) log-log plots of $R_g^2$ vs. $N$. The error bars denote the standard errors.} 
\label{fig-4}
\end{figure}
In Fig. \ref{fig-4}(a), we show $R_g^2$ vs. $\kappa$ for $\kappa\!=\!0$ and $\kappa\!=\!0.1$. To see $D_f$ for $\kappa\!\to\! 0$, we plot $R_g^2$ vs. $N$ in Fig. \ref{fig-4}(b) in a log-log scale. We draw the straight lines by fitting the data to $R_g^2\!\sim\! N^{2/D_f}$ in Eq. (\ref{size-exponents}), and we have
\begin{eqnarray}
\label{H_fluid}
&&D_f=2.08\pm0.34 \quad (b\!=\!0),\nonumber \\
&&D_f=2.06\pm0.41 \quad (b\!=\!0.1). 
\end{eqnarray}
Relatively large errors of $D_f$ for $\kappa\!\to\! 0$ come from the large errors of the data $R_g^2$, and the the latter come from the large fluctuations of surface shape in the BP phase. The values of $D_f$ for both $\kappa\!=\!0$ and $\kappa\!=\!0.1$ are consistent with the expectation $D_f\!=\!2$ for $\kappa\!\to\! 0$. We note that the fractal dimension of the random walks is $D_f\!=\!2$, and that the random walk can be described by 
the Gaussian bond potential $S_1$ for an ideal chain. On the other hand, we have $D_f\!=\!5/3$ for the SA random walk, which is a model of linear polymer \cite{Doi-Edwards-1986}. Thus, the SA fluid surface at $\kappa\!\to\!0$ and ${\it \Delta p}\!\to\!0$ is close to the random walk.

We should note that the BP surface is characterized by a density of vertices $\rho(r)\!\sim\!1/r$, where $r\!=\!0$ is the center of mass of the surface, and $\rho(r)$ is the total number of vertices in the unit volume at $r$. Actually, it is possible to assume $\rho(r)\sim r^d (d\geq -2)$ for simplicity. Then, we have $\langle R_g^2\rangle\!:=\!\int r^2 \rho dV/\int \rho dV\!\sim\! R^2$, where $R$ is the radius of the minimal sphere in which the BP surface is included. Therefore, we have $d\!=\!-1$ from the assumption that $N\!:=\! \int \rho dV\sim R^{d+3}$, and we have $D_f\!=\!2$. The implication of $\rho(r)\!\sim\!1/r$ means that the total number $dN$ of vertices inside $dV\!=\!4\pi r^2dr$ is given by $dN \!\sim\!r dr$.

We compare the results $D_f$ in Eq. (\ref{H_fluid}) at $\kappa\simeq 0$ with $D_f^{b.p.}\!=\!1.93\pm0.06$ in Ref. \cite{KOIB-PRE-2003}, where the model is a phantom and fluid, which is defined by the Gaussian bond potential $S_1$ and the deficit angle energy $S_3$. Here $S$ equals $S_1\!+\!\alpha S_3$ with the curvature coefficient $\alpha$. The BP phase in the model of Ref. \cite{KOIB-PRE-2003} appears only in an intermediate range of $\alpha$; the surface is collapsed both at $\alpha\!\to\! 0$ and at $\alpha\!\to\! \infty$. Since $D_f^{b.p.}\!=\!1.93\pm0.06$ in Ref. \cite{KOIB-PRE-2003} is comparable to $D_f$ at $\kappa\simeq 0$ of the model in this paper, the BP phase observed in fluid surface models is universal in the sense that it is independent of the model and even whether the model has a SA interaction or not. This is in sharp contrast to the case of linear chain model mentioned above.  

\begin{figure}[htb]
\centering
\includegraphics[width=11cm]{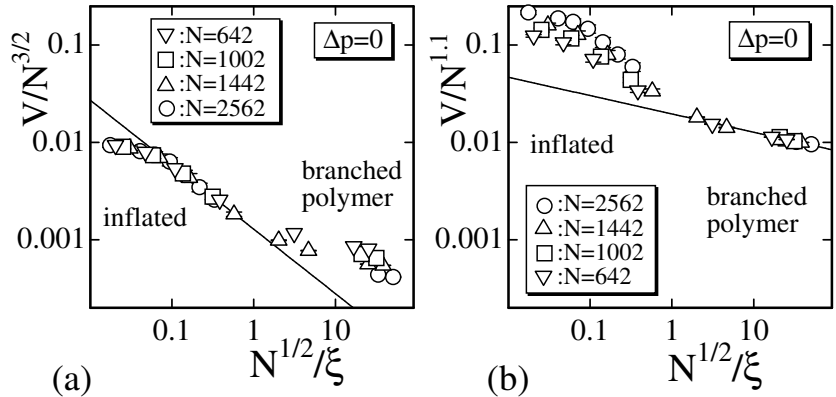}
\caption{ Two different scaling behaviors (a) $V/N^{3/2}$ vs. $N^{1/2}/\xi$ and  (b)  $V/N^{1.1}$ vs. $N^{1/2}/\xi$, where $\xi\!=\!\exp(4\pi\kappa/3)$ is the persistence length.} 
\label{fig-5}
\end{figure}

Nextly, we show the scaling behavior for the enclosed volume $V$ such as in \cite{Gompper-Kroll-PRE1995}:
\begin{equation}
\label{V-scaling}
V\sim N^{2\nu}X(y),\quad y=\sqrt{N}/\xi
\end{equation}
where $\xi\!=\!a\exp(4\pi\kappa/3)$; $a\!=\!1$ is the persistence length. 

We see from Fig. \ref{fig-5}(a) that $V$ scales according to Eq. (\ref{V-scaling}) for $2\nu\!=\!3/2$ in the transition region close to $\kappa\!=\!1.5$. This behavior is consistent with that for small $\kappa$ under ${\it \Delta p}\!=\!0$ in Ref.  \cite{Gompper-Kroll-PRE1995}. The slope $\sigma$ of the straight line is $\sigma\!=\!-0.65\pm0.04$, which gives $\nu_V\!=\!1.18$ for $V\!\sim N^{\nu_V}$ in Eq. (\ref{size-exponents}) and is consistent with $\nu_V$ for $\kappa\!=\!1.3$ in Table \ref{table-1} in the transition region.   

 It is also confirmed from Fig. \ref{fig-5}(b) that $V$ scales according to Eq. (\ref{V-scaling}) for $2\nu\!=\!1.1$ in the region of sufficiently small $\kappa$ including $\kappa\!=\!0$. This behavior was not found in the model of  Ref. \cite{Gompper-Kroll-PRE1995}, where the simulation was not performed in the sufficiently small $\kappa$ region for  ${\it \Delta p}\!=\!0$.  The slope $\sigma$ of the straight line in Fig. \ref{fig-5}(b) is $\sigma\!=\!-0.17\pm0.01$, which gives $\nu_V\!=\!1.02$ for $V\!\sim N^{\nu_V}$ in Eq. (\ref{size-exponents}), and this $\nu_V\!=\!1.02$ is close to $\nu_V\!=\!1$ which is expected in the BP phase. The jump in the scaling of $V/N^{2\nu}$ vs. $\sqrt{N}/\xi$ supports that the inflated phase is separated from the BP phase by a second order phase transition. 

\begin{figure}[htb]
\centering
\includegraphics[width=10cm]{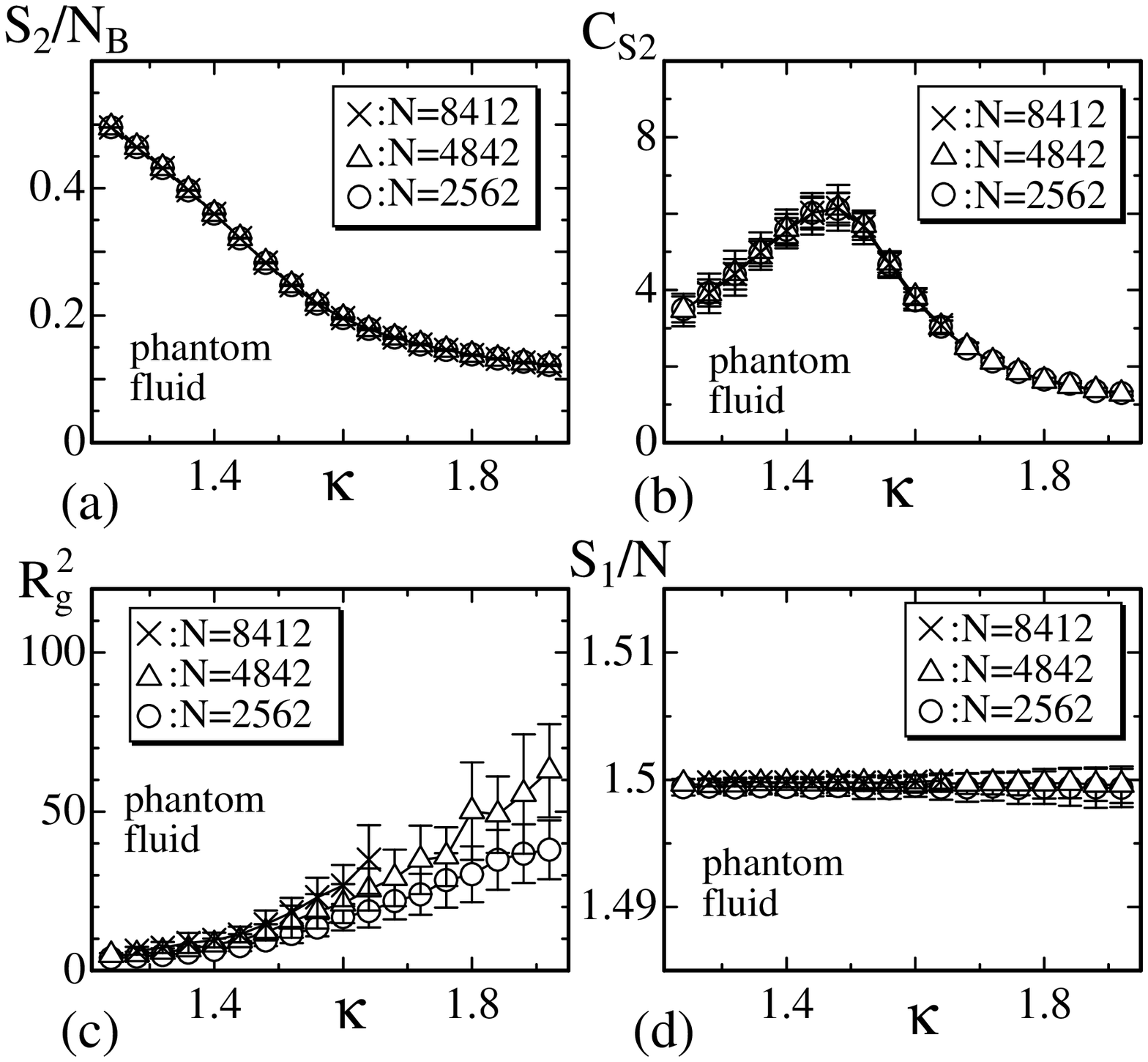}
\caption{ The results of phantom fluid model: (a) The bending energy $S_2/N_B$ vs. $\kappa$, (b) the specific heat $C_{S_2}$ vs. $\kappa$, (c) the mean square gyration $R_g^2$ vs. $\kappa$ and (d) the Gaussian bond potential $S_1/N$ vs. $\kappa$. } 
\label{fig-6}
\end{figure}
Finally in this subsection, we compare results of the SA model with those of the phantom fluid model. The phantom model is defined by the Hamiltonian $S\!=\!S_1\!+\!\kappa S_2$, which differs from $S$ in Eq. (\ref{Disc-Eneg}) by $-{\it \Delta}pV\!+\!U$. The partition function of the phantom model is identical to the one in Eq. (\ref{Part-Func-2}). The coordination number $q_i$, which is the total number of bonds emanating from the vertex $i$, is limited such that $3\!\leq\! q_i\!\leq\!40$ for all $i$ in the phantom model.  The total number of simulations is $5\!\times\! 10^8\sim 8\!\times\! 10^8$ MCS for the surfaces with $N\!=\!2562$, $N\!=\!4842$ and  $N\!=\!8412$ after sufficiently large thermalization MCS. 

From the bending energy $S_2/N_B$ and the specific heat $C_{S_2}\!=\!(\kappa^2/N)\langle \left(S_2\!-\!\langle S_2\rangle \right)^2 \rangle$ plotted in Figs. \ref{fig-6}(a),(b), we see that the model has no crumpling transition. In fact, the peak value of  $C_{S_2}$ remains unchanged with increasing $N$. This is in sharp contrast with the results of the phantom tethered model \cite{KOIB-PRE-2005}. Indeed, the crumpling transition can be seen in the phantom tethered model. It accompanies the first-order transition of surface fluctuations, which is reflected in the discontinuous change of $S_2/N_B$ \cite{KOIB-PRE-2005}. The mean square radius of gyration $R_g^2$ appears to change non-smoothly or randomly (Fig. \ref{fig-6}(c)). This non-smooth change of $R_g^2$ is due to the self-intersections of surface. Moreover, no inflated surface is seen in the region $\kappa \!\leq\!1.92$ at least. The fact that no inflated phase appears in the region $\kappa\!\simeq\! 1.9$ is in a sharp contrast with the case of the SA model, where the inflated phase is seen in the region  $ \kappa \!>\!1.5$ for ${\it \Delta p}\!=\!0$ at least. The Gaussian bond potential $S_1/N$ shows the expected behavior: $S_1/N\!=\!1.5$  (Fig. \ref{fig-6}(d)), which comes from the scale invariant property of the partition function \cite{WHEATER-JP1994}.  

We should note that the phase structure of the fluid model is expected to depend on the coordination number. The dependent integration measure equals $\Pi_i d{\bf r}_i q_i^\alpha$ with $\alpha\!=\!3/2$ both on the SA and phantom surfaces \cite{Gompper-Kroll-PRE1995,David-NP-1985,DAVID-SMMS2004}. In this paper, $\alpha$ is fixed to $\alpha\!=\!0$ in both models.  

\subsection{Fixed bending rigidity $\kappa=1.5$}
In this subsection, we vary ${\it \Delta p}$ in the neighborhood of ${\it \Delta p}\!=\!0$ by fixing the bending rigidity to $\kappa\!=\!1.5$. In this region $C_V$ has a peak $C_V^{\rm max}$ on the surface with $642\!\leq\! N\!\leq\! 2562$ at least as we have seen above. 

\begin{figure}[tb]
\centering
\includegraphics[width=10cm]{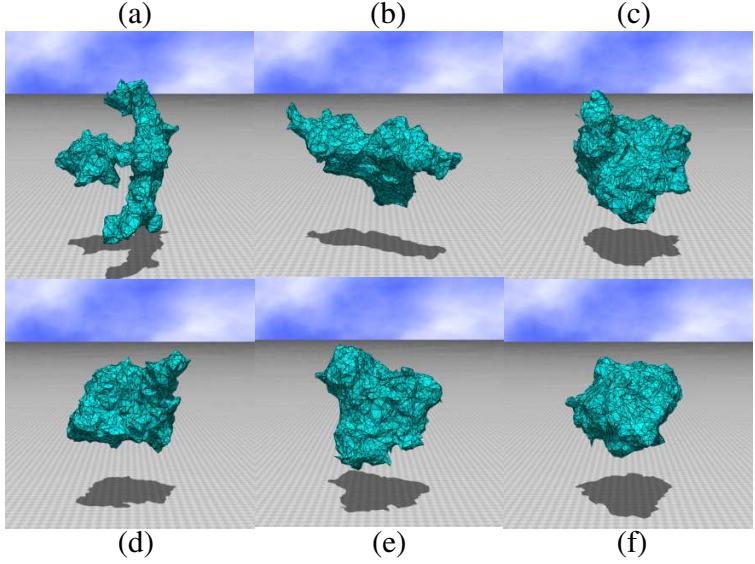}
\caption{(Color online)  Snapshot of surfaces obtained at (a) ${\it \Delta p}\!=\!-0.04$, (b) ${\it \Delta p}\!=\!-0.02$, (c) ${\it \Delta p}\!=\!-0.01$, (d) ${\it \Delta p}\!=\!0$, (e) ${\it \Delta p}\!=\!0.02$, and (f) ${\it \Delta p}\!=\!0.04$.  The bending rigidity is fixed to $\kappa\!=\!1.5$. The scales of the figures are the same.  } 
\label{fig-7}
\end{figure}
We see from Figs. \ref{fig-7}(a)-(f) that the surface is in the BP (inflated) phase for ${\it \Delta p}\!<\!0$ (${\it \Delta p}\!>\!0$). Thus, the expected transition is the one which can also be expected when ${\it \Delta p}\!\to\!0$ in the fluid vesicle model in \cite{Gompper-Kroll-PRE1995}.  

\begin{figure}[hbt]
\centering
\includegraphics[width=11cm]{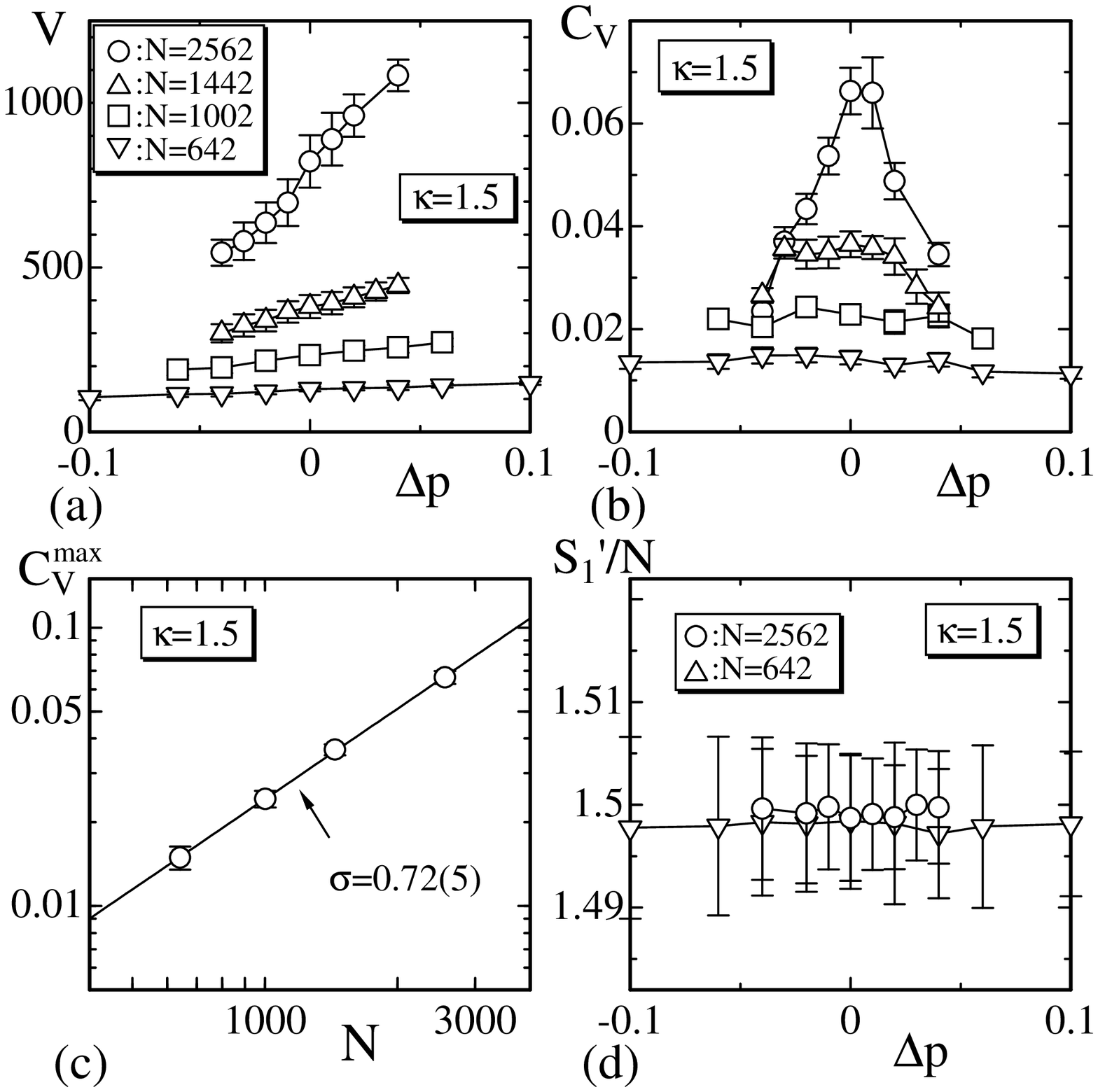}
\caption{ (a) The enclosed volume $V$ vs. $\kappa$ for $\kappa\!=\!1.5$, (b) the variance $C_V$ vs. $\kappa$, (c) log-log plots of $C_V^{\rm max}$ vs. $N$, and (d) $S_1^\prime/N$ vs. $\kappa$. } 
\label{fig-8}
\end{figure}
The enclosed volume $V$ rapidly varies in the neighborhood of ${\it \Delta p}\!\simeq\!0$ when $N$ grows (Fig. \ref{fig-8}(a)). This indicates the existence of phase transition between the inflated and BP phases. The variance $C_V$ has an expected peak, which also grows with the increase of $N$ (Fig. \ref{fig-8}(b)).  

 In order to see the scaling behavior, we plot the peak value $C_V^{\rm max}$ vs. $N$ in Fig. \ref{fig-8}(c) in a log-log scale. The lattice size $N\!=\!2562$ is not so large compared to that of the phantom surface simulations \cite{KD-PRE2002,KOIB-PRE-2005}. However, we suppose that the lattice size is large enough to see the following scaling:
\begin{equation}
\label{CV-scaling-fluid}
C_{V}^{\rm max}\sim N^{(3/2)\sigma}, \quad \sigma=0.72\pm 0.05.
\end{equation}
The straight line in Fig. \ref{fig-8}(c) is drawn by the least-squares fitting of the data to the relation ({\ref{CV-scaling-fluid}}). From the finite-size scaling theory and from the fact that $\sigma < 1$ we conclude that it is the second order transition. 

The phase transition is called the first (second) order if the first-order (second-order) derivative of $Z$ has a gap.  The variance $C_V$ is given by the second-order derivative of $Z$ with respect to ${\it \Delta p}$. Therefore, the result shown in Eq. (\ref{CV-scaling-fluid}) confirms that the transition from the BP to the inflated phase is a true continuous transition.

From the scale invariance of $Z$, we have 
\begin{equation}
S_1^\prime=S_1-(3/2) {\it \Delta p} V=(3/2)N, \quad (N\to\infty)
\end{equation}
in the limit of $N\to\infty$. Thus, we see that  $S_1^\prime/N\!=\!1.5$ is satisfied in the MC data (Fig. \ref{fig-8}(d)). This implies that the SA interaction is correctly implemented in the MC simulations. 

We should note that the continuous transition is not reflected in the specific heat $C_{S_2}$ for the bending energy $S_2$ defined by $C_{S_2}\!=\!(\kappa^2/N) \langle \left(S_2\!-\!\langle S_2\rangle \right)^2 \rangle$. In fact,  a very small peak can only be seen in $C_{S_2}$ at $\kappa\!=\!1.5$ on the $N\!=\!2562$ surface. This is in sharp contrast with the phantom tethered model, which has the first-order transition characterized by discontinuous change of $S_2$ \cite{KOIB-PRE-2005}. The reason for this difference is that the first-order transition in the phantom model separates the inflated phase from the crumpled (or collapsed) phase, while the continuous transition in the SA model separates the inflated phase from the BP phase; there is no collapsed phase in the SA model. However, the possibility that the peak in $C_{S_2}$ grows larger and larger on larger surfaces is not completely eliminated. Indeed, a very low momentum mode of surface fluctuation is probably connected with the surface fluctuations of the SA fluid surface just like in the case of the crumpling transition on the fixed-connectivity phantom surfaces, where the first-order transition is seen only on the lattices  of size $N\geq 7000$ approximately \cite{KOIB-PRE-2005}.

\begin{table}[tb]
\caption{ The size exponents $\nu_V$, ${\bar \nu}_V$, and $\nu_{R^2}(=2/D_f)$ obtained under $\kappa\!=\!1.5$.}
\label{table-2}
\begin{center}
 \begin{tabular}{ccccccccc}
 \hline
 ${\it \Delta p}$ && $\nu_V$ && ${\bar \nu}_V$ && $\nu_{R^2}$ && $D_f$     \\
 \hline
 $-0.04$  && $1.12\pm0.07$ && $0.75\pm0.05$ && $1.03\pm0.12$ && $1.93\pm0.23$ \\
 $-0.02$  && $1.20\pm0.08$ && $0.80\pm0.05$ && $0.94\pm0.08$ && $2.14\pm0.18$ \\
 $0$     && $1.33\pm0.07$ && $0.89\pm0.05$ && $0.92\pm0.05$ && $2.17\pm0.13$ \\
 $0.02$ && $1.42\pm0.06$ && $0.95\pm0.04$ && $0.92\pm0.03$ && $2.16\pm0.08$ \\
 $0.04$ && $1.51\pm0.05$ && $1.00\pm0.03$ && $0.93\pm0.03$ && $2.15\pm0.08$ \\
 \hline
 \end{tabular} 
\end{center}
\end{table}
The size exponents are shown in Table \ref{table-2}. There the data for ${\it \Delta p}\!=\!0$ are the same as those for $\kappa\!=\!1.5$ in Table \ref{table-1}. The exponent $\nu_V\!=\!1.51(5)$ for ${\it \Delta p}\!=\!0.04$ is almost exactly identical with $\nu_V\!=\!1.5$ expected on the inflated surface, and this indicates that the surface is really inflated for ${\it \Delta p}\!=\!0.04$. When ${\it \Delta p}$ decreases to ${\it \Delta p}\!=\!-0.04$, we have $\nu_V\!=\!1.12(7)$, which is slightly larger than $\nu_V\!=\!1$. This reflects that the BP surface for ${\it \Delta p}\!=\!-0.04$ is inflated a little (Fig. \ref{fig-7}(a)).  At the transition point ${\it \Delta p}\!=\!0$ we expect ${\bar \nu}_V\!=\!\nu_{R^2}$, where the surface is inflated, and indeed this equality is satisfied (${\bar \nu}_V\!=0.89(5)$, $\nu_{R^2}\!=\!0.92(5)$) within the error bounds.

\begin{figure}[htb]
\centering
\includegraphics[width=11cm]{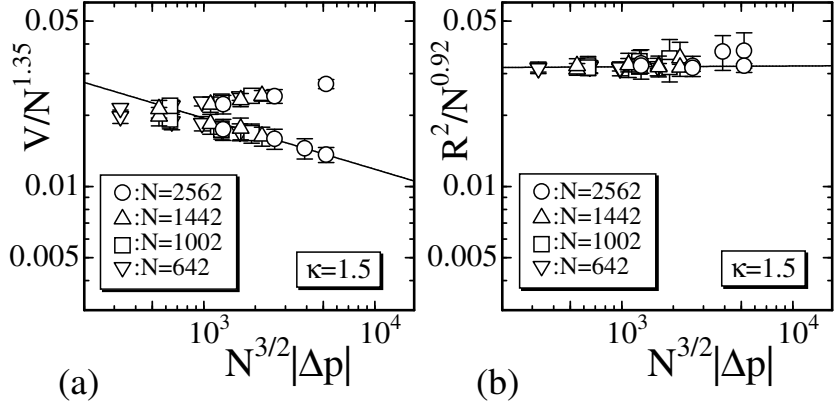}
\caption{ Two different scaling behaviors $V/N^{1.35}$ vs. $N^{3/2}|{\it \Delta}p|$ corresponding to  ${\it \Delta}p>0$ and  ${\it \Delta}p<0$.} 
\label{fig-9}
\end{figure}
The enclosed volume $V$ is expected to scale according to \cite{Leibler-etal-PRL1987}
\begin{equation}
\label{V-dp-scaling}
V\sim N^{2\nu}Y(y),\quad y=N^{3/2}\left|{\it \Delta}p\right|.
\end{equation}
We find from Fig. \ref{fig-9}(a) that $V/N^{2\nu} (2\nu\!\simeq\! 1.35)$ has two branches as expected, although the range $-0.04\!\leq\!{\it \Delta}p|\!\leq\!0.04$ is not so wide. The value of $2\nu\!\simeq\! 1.35$ is almost identical to  $\nu_V\!=\! 1.33(7)$  in Table \ref{table-2} at ${\it \Delta p}\!=\!0$. 
The slope $\sigma$ of the straight line is $\sigma=-0.22\pm0.09$, which describes the scaling property in the BP phase for small negative ${\it \Delta}p$. The mean square radius of gyration $R_g^2$ is expected to satisfy $R_g^2\sim N^{2\nu}Z(y),\; y=N^{3/2}\left|{\it \Delta}p\right|$. We find that $2\nu\!=\!0.92$, which is equal to $\nu_{R^2}$ in Table \ref{table-2} at ${\it \Delta p}\!=\!0$. For  $R_g^2$ in Fig. \ref{fig-9}(b) the straight line is drawn  using the data at ${\it \Delta}p\!<\!0$ in the BP phase; the slope equals $0.007\pm0.01$. These results indicate that $V/(R_g^2)^{3/2}\to 0\; (N\to\infty)$ in the BP phase for small negative ${\it \Delta p}$ just like in the model of the ring polymer for negative ${\it \Delta p}$ on the plane \cite{Leibler-etal-PRL1987}. 

\subsection{Phase diagram}
\begin{figure}[htb]
\centering
\includegraphics[width=11cm]{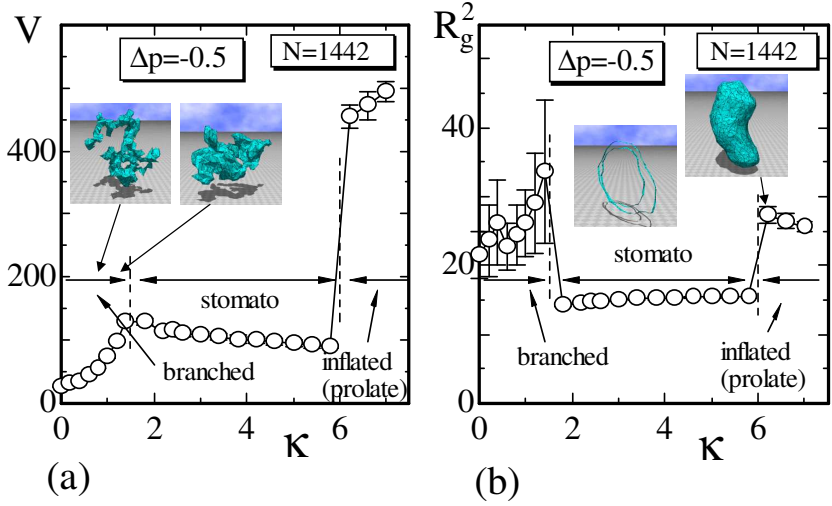}
\caption{ (Color online) (a) The enclosed volume $V$ vs. $\kappa$, and (b) the mean square gyration  $R_g^2$ vs. $\kappa$, under  ${\it \Delta}p\!=\!-0.5$. Snapshots of surface and the surface section (stomatocyte) are shown.} 
\label{fig-10}
\end{figure}
In this subsection we present a tentative $\kappa$ vs. ${\it \Delta p}$ phase diagram . First of all, in Figs. \ref{fig-10} (a),(b) we compare the graphs of $V$ vs. $\kappa$ and $R_g^2$ vs. $\kappa$ obtained for ${\it \Delta p}\!=\!-0.5$ with snapshots of surfaces. We see three different phases: BP, stomatocyte and inflated (or prolate). The latter two are expected to be separated by a first-order transition because $V$ changes discontinuously at the phase boundary. To the contrary, the first two - the BP and the stomatocyte - are separated by a continuous transition because $V$ changes continuously, although $R_g^2$ exhibits a break at the transition point.

The phase transition is of the first order if there exists a physical quantity which changes discontinuously at the transition point. Thus the transition between the BP and stomatocyte phases is a first-order one. The bending energy $S_2/N_B$, which is not shown in the figure, also changes continuously at this phase boundary. These three phases can also be seen on the $N\!=\!2562$ surface when $\kappa$ varies in the region $1.4\!<\!\kappa\!<\!3.6 $ under  ${\it \Delta p}\!=\!-0.3$  and  ${\it \Delta p}\!=\!-0.2$.

\begin{figure}[htb]
\centering
\includegraphics[width=10cm]{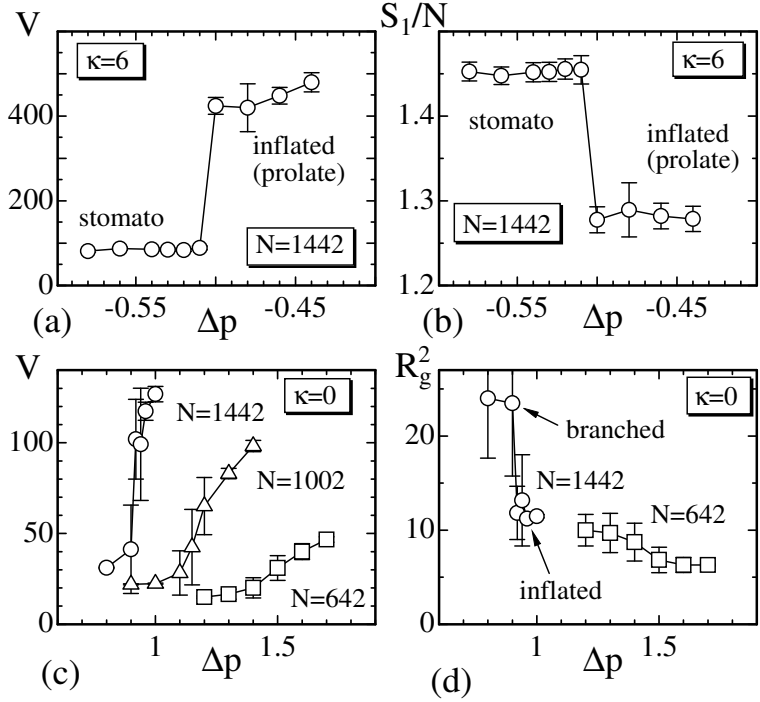}
\caption{ (a) The enclosed volume $V$ vs. ${\it \Delta}p$, (b) the Gaussian bond potential $S_1/N_B$ vs.  ${\it \Delta}p$ at $\kappa\!=\!6$, and  (c) $V$ vs. ${\it \Delta}p$, (d) the mean square gyration  $R_g^2$ vs.  ${\it \Delta}p$ at $\kappa\!=\!0$. }  
\label{fig-11}
\end{figure}
The discontinuous changes in the enclosed volume $V$ and Gaussian bond potential $S_1/N$ can also be seen by varying ${\it \Delta p}$ when $\kappa$ is fixed to a relatively large value such as $\kappa\!=\!6$ (Figs. \ref{fig-11} (a), (b)). On the $N\!=\!1442$ surface, the transition point on the $\kappa$ axis is located at  ${\it \Delta p}\!\simeq\! -0.5$, where the stomatecyte (${\it \Delta p}\!<\! -0.5$)  and inflated (${\it \Delta p}\!>\! -0.5$)  phases are separated. In the inflated phase at ${\it \Delta p}\!\simeq\! -0.5$, the surface is almost tubular or prolate.  On the axis $\kappa\!=\!0$, the BP and inflated phases are separated by the first-order transition (Figs. \ref{fig-11} (c), (d)). The surface in the inflated phases at the transition point is very rough. This is because the surface has no bending elasticity when $\kappa\!=\!0$. At this transition, the inflated surface collapses into the BP surface accompanying a discontinuous change of $V$ (Figs. \ref{fig-11} (c)). The transition point ${\it \Delta}p_c$ moves left on the ${\it \Delta}p$ axis  as $N$ increases for $\kappa\!=\!0$. By extrapolating ${\it \Delta}p_c(N)$ linearly against $1/N$, we have  ${\it \Delta}p_c\!=\!0.4\!\sim\! 0.5$ in the limit of $N\!\to\!\infty$.

\begin{figure}[htb]
\centering
\includegraphics[width=11.0cm]{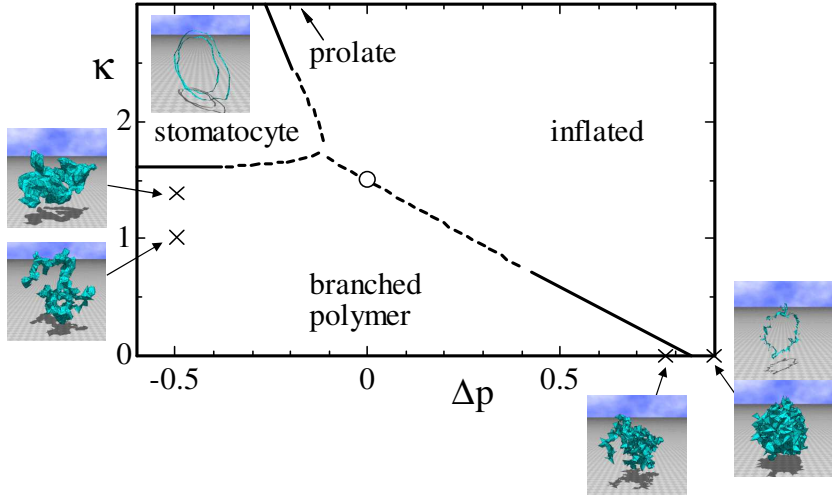}
\caption{(Color online)  An expected phase diagram close to the point $\kappa\!=\!1.5, {\it \Delta}p\!=\!0$ ($\circ$). The solid (dashed) line represents a first (second) order phase transition. These lines are obtained from the MC surface simulation results for the $N\!=\!1442$ and $N\!=\!2562$ surfaces. The location of triply duplicated critical point is only tentative, and the points where the second order transition changes to the first order one are also tentative.} 
\label{fig-12}
\end{figure}
So in Fig. \ref{fig-12} we demonstrate a preliminary phase diagram. The solid (dashed) line represents a first (second) order transition. We have a triply duplicated point close to $\kappa\!=\!1.5, {\it \Delta}p\!=\!0$; the location of the point is unclear in Fig. \ref{fig-12}. These three phases are expected to be separated by continuous transitions in the region close to the tricritical point. The volume $V$ discontinuously changes between the stomatocyte and prolate phases (see Fig. \ref{fig-10}(a)), and it changes in the same way between the BP and inflated phases at $\kappa\!=\!0$ (Figs. \ref{fig-11}(c),(d)). In the limit of $N\!\to\!\infty$, a first order transition separates the BP from inflated phases at ${\it \Delta}p_c\!=\!0.4\!\sim\! 0.5$ as mentioned above on the line of $\kappa\!=\!0$. The solid line is drawn using the data of $N\!=\!1442$ surface, and hence ${\it \Delta}p_c$ is located at relatively larger value of ${\it \Delta}p$ than ${\it \Delta}p_c(N\!\to\!\infty)$ on the axis. To the contrary, $V$ continuously changes at the phase boundary between the stomatocyte and BP phases at least in the region ${\it \Delta}p\geq\!-0.5$ as we see in Fig. \ref{fig-10}(a).

\section{Summary and conclusions}
Using the canonical Monte Carlo simulation technique, we have studied a self-avoiding surface model on dynamically triangulated fluid lattice with sphere topology and of size up to $N\!=\!2562$. The Hamiltonian includes the pressure term $-{\it \Delta p}V$, where ${\it \Delta p}$ is the pressure difference between the inner and outer sides of the surface, and $V$ is the enclosed volume. 

We find that the model undergoes a continuous transition when ${\it \Delta p}\!=\!0$ and $\kappa\!=\!1.5$. This transition  separates the branched polymer phase for $\kappa\!\to\!0$ from the inflated phase for sufficiently large $\kappa$ when ${\it \Delta p}\!=\!0$. At the transition we observe a scaling property of the variance $C_V$ such that $C_V^{\rm max}\sim N^{(3/2)\sigma} (\sigma\!=\! 0.72)$. Moreover, the jump in the scaling of $V$ in the small bending region when ${\it \Delta p}\!=\!0$ supports the occurrence of the phase transition. We also demonstrate that no surface fluctuation transition accompanies the observed transformation at least up to the number of the surface vertices $N\!=\! 2562$. 

%
\vspace*{3mm}
\noindent
{\bf Acknowledgment}\\

This work is supported in part by a Promotion of Joint Research, Nagaoka University of Technology. We are grateful to Koichi Takimoto in Nagaoka University of Technology for the support. We also acknowledge Kohei Onose for the computer analyses.


\end{document}